\documentclass[11pt,a4paper]{article}
\pdfoutput=1
\usepackage{jheppub}
\usepackage[sort&compress]{natbib}
\usepackage{url}
\usepackage{hyperref}
\usepackage{amsfonts}% Include figure files
\usepackage{epsfig}\usepackage{dcolumn}% Align table columns on decimal point
\usepackage{bm}% bold math
\usepackage{slashed}
\graphicspath{{./figuras/}}
\usepackage{subfigure}

\newcommand{\gev}{\; \hbox{GeV}}
\newcommand{\ifb}{\; \hbox{fb}^{-1}}
\newcommand{\zp}{Z^{\prime}}

\newcommand{\mzp}{M_{Z'}}

\newcommand{\gx}{g_{\chi}}

\title{ {\color{blue} \mbox{Dark Matter Complementarity and the Z$^\prime$ Portal}}}
\author{Alexandre Alves$^a$,}
\author{Asher Berlin$^{b,c}$,}
\author{Stefano Profumo$^d$,}
\author{Farinaldo S. Queiroz$^{d,e}$}

\affiliation{$^a$Departamento de Ci\^encias Exatas e da Terra,
Universidade Federal de S\~ao Paulo, Diadema-SP, 09972-270, Brazil}
\affiliation{$^b$Enrico Fermi Institute, University of Chicago, Chicago, IL 60637}
\affiliation{$^c$Kavli Institute for Cosmological Physics, University of Chicago, Chicago, IL 60637}
\affiliation{$^d$Department of Physics and Santa Cruz Institute for Particle Physics
University of California, Santa Cruz, CA 95064, USA}
\affiliation{$^e$ International Institute of Physics, UFRN, Av. Odilon Gomes de Lima, 1722 - Capim Macio - 59078-400 - Natal-RN, Brazil}

\emailAdd{aalves@unifesp.br}
\emailAdd{berlin@uchicago.edu}
\emailAdd{profumo@ucsc.edu}
\emailAdd{fdasilva@ucsc.edu}

\abstract{$\zp$ gauge bosons arise in many particle physics models as mediators between the dark and visible sectors. We exploit dark matter complementarity and derive stringent and robust collider, direct and indirect constraints, as well as limits from the muon magnetic moment. We rule out almost the entire region of the parameter space that yields the right dark matter thermal relic abundance, using a generic parametrization of the $Z^{\prime}$-fermion couplings normalized to the Standard Model Z-fermion couplings for dark matter masses in the 8 GeV-5 TeV range. We conclude that mediators { lighter than $2.1$~TeV are excluded regardless of the DM mass, and that depending on the $\zp-fermion$ coupling strength much heavier masses are needed to reproduce the DM thermal relic abundance while avoiding existing limits}.
}
    
\begin{document} 
\maketitle
\flushbottom
    
%%%%%%%%%%%%%%%%%%%%%%%%%%%%%%%%%%%%%%%%%%%%%%%%%%%%%%%%%%%
\section{Introduction}

One of the most exciting and tantalizing puzzles of modern cosmology and particle physics lies with the nature of dark matter (DM). DM comprises roughly 23\% of the energy budget of the Universe and its existence has been so far inferred only from gravitational effects. The most compelling particle DM candidates are the so called Weakly Interacting Massive
Particles (WIMPs), arising in a wide variety of
well-motivated theories beyond the Standard Model (SM) such as supersymmetry and models with extra space dimensions. WIMPs can naturally account for the observed DM abundance in the framework of standard thermal freeze-out, and they are often within reach of current and future experiments (see Ref.~\cite{Baer:2014eja,Strigari:2013iaa,Bertone:2004pz} for extensive overviews on particle
DM).

Although the presence of DM has been ascertained only gravitationally, there are several observations which indicate that the direct detection of DM particles is around the corner. Direct detection experiments attempt to measure the recoil energy deposited by DM particles in underground experiments. The measured rate of scattering events, after the subtraction of background, singles out a preferred region in terms of the scattering cross section and DM mass (see Ref.~\cite{dirdet} for a review of current status and prospects in this field). %Some direct detection experiments have reported excess events consistent with $7-30$~GeV WIMPs scattering off of nuclei with a scattering cross section of $10^{-(41-42)}~{\rm cm^2}$, such as COGENT, DAMA, CREST, and CDMS-Si \cite{Kelso:2011gd1,Kelso:2011gd2}. However, numerous other experimental collaborations, including XENON and LUX, have not confirmed any such excesses \cite{Agnese:2013jaa}, excluding the majority of the parameter space (cross section $\times$ mass) that is best fit by the DM interpretation. Furthermore, attempts to reconcile these findings by invoking isopin violation to suppress the XENON/LUX bounds no longer suffice. In Ref.~\cite{Va'vra:2014mma} 1 GeV WIMPs were postulated to circumvent the null results and explain DAMA modulation in the presence of OH impurities in the DAMA detector. However, Ref.~\cite{Profumo:2014mpa} has proved otherwise. In summary, there is no conclusive evidence for DM scattering from direct detection data.

Indirect DM detection consists of the observation of DM annihilation products by satellite or ground-based telescopes. This is a promising and complementary avenue, since it extracts information concerning the DM distribution and annihilation cross section \cite{complementarity}. Ref.~\cite{indirdet} gives an account of the current status of indirect searches and of future observational avenues.

Concerning collider searches, the typical signature of DM production is the presence of
missing energy, since WIMPs simply escape the detector. Hence, collider searches for DM production typically comprise jets and missing energy and provide rather complementarity limits \cite{An:2012ue}. However, often times the most stringent limits do not come from missing energy signatures, but instead from the production of SM particles with the mediator being produced in the s-channel. Looking for resonances in dileptons or dijets has proven to be a prominent way to constrain dark sectors. As we shall discuss further, those bounds are more stringent than the direct and indirect DM detection ones if the mediators interact with leptons. For recent leptophilic DM models see Refs.\cite{Kopp:2014tsa1,Kopp:2014tsa2,Kopp:2014tsa3}.

Several Dirac fermion DM models have been proposed, with DM particle masses in the GeV-TeV range, where the DM is coupled to the visible sector via $\zp$ gauge bosons, arising from a $U(1)_X$ gauge sector. Such gauge bosons have been extensively studied in many different contexts \cite{An:2012ue,Kopp:2014tsa1,Arcadi:2014lta}, but our focus here is on the dark matter phenomenology.

Instead of choosing a particular model, we perform here a comprehensive, model-independent analysis. Aside from an overall rescaling, we assume that the $\zp$ has similar couplings to SM $Z$, an assumption studied in the previous literature \cite{An:2012ue,Alves:2013tqa,Arcadi:2014lta}. By studying regimes in which the $\zp$-SM fermion couplings are suppressed by an overall factor, we implicitly encompass several SM extensions such as sequential $\zp$ models \cite{Barger:1980dx}, 3-3-1 models \cite{Profumo:2013sca1,Martinez:2014lta}, E-6 models \cite{Nie:2001ti}, Left-Right \cite{Wong:1992qa}, etc. We present results for a broad range of particle DM masses, specifically: 8 GeV, 15 GeV, 50 GeV, 500 GeV, 1 TeV, and 5 TeV, with $\zp-\chi-\chi$ couplings ranging from $10^{-5}-10$.

Due to the presence of a spin 1 mediator that couples to the muon, important constraints potentially arise from contributions to the muon anomalous magnetic moment. We use here a new public code for computing the muon magnetic moment to derive limits on the mediator mass described in Ref.~\cite{Queiroz:2014zfa}.  Since only the muon's interactions with the $\zp$ enter into this calculation, the muon magnetic moment offers quite interesting bounds in the $\zp$-DM suppressed couplings regime where direct and indirect detection experiments lose sensitivity. 

As mentioned, extensive literature exists on the subject topic we focus on here. Our work is similar in vein to that presented in Ref.~\cite{An:2012ue,Kopp:2014tsa1,Kopp:2014tsa2,Kopp:2014tsa3,Profumo:2013sca1,Profumo:2013sca2,Profumo:2013sca3,
Profumo:2013sca4,Profumo:2013sca7,Profumo:2013sca5,Profumo:2013sca6}. Ref.\cite{An:2012ue} focused on the leptophobic regime, whereas Refs. \cite{Kopp:2014tsa1,Kopp:2014tsa2,Kopp:2014tsa3} discussed the leptophilic scenario. In Refs.\cite{Profumo:2013sca1,Profumo:2013sca2,Profumo:2013sca6} the DM phenomenology was studied in a particular model, and Ref.\cite{Profumo:2013sca3} focused on resonances scenarios. Ref.\cite{Profumo:2013sca4} performed a similar analysis to ours with an older data,  without computing indirect DM detection and g-2 bounds, and assumed  $\zp-{\rm up\, quarks}$ and $\zp-{\rm down\, quarks}$ to be the same, whereas in Ref.\cite{Profumo:2013sca7} the kinetic mixing window has been explored. In Ref.\cite{Profumo:2013sca5} the authors discussed the very light mediator regime only.

The present study, however, differs from the mentioned previous studies in several key ways:

\begin{itemize}
\item We use a generic parametrization to describe the Dirac fermion DM - SM fermion interactions in the context of the $Z^{\prime}$ portal.

\item We derive up to date spin-independent and spin-dependent limits using LUX and XENON100 data.

\item We derive indirect detection limits, including the important annihilation channel to on-shell $\zp$s when $m_{\zp} < m_{\chi}$.

\item We  carry out a detailed and comprehensive collider analysis, using up to date dilepton limits.

\item We obtain limits from the muon magnetic moment on the $Z^{\prime}$ mass. 
\end{itemize}

We begin our study with an outline of the $\zp$ DM portal we focus on.
  
\section{Z$^{\prime}$ Portal Dark Matter Model}
The existence of new gauge symmetries is among the best motivated extensions to the SM. In particular, in the low energy limit of many new-physics scenarios, an additional heavy electrically neutral gauge boson, $\zp$, generically possesses sizable couplings to SM quarks and leptons~\cite{Langacker:2008yv}. DM that interacts with the SM purely through electroweak interactions is strongly constrained~\cite{deSimone:2014pda}, and therefore a dark sector that communicates with the SM via a new $\zp$ remains a very natural scenario. In this paper, we study a generic $\zp$ model, and parametrize the $\zp$-fermion couplings so that the $\zp$ couples similarly to the SM $Z$ boson (see e.g. Ref.~\cite{Arcadi:2014lta}). This regime is also known as a sequential $\zp$. In this case, the crucial differences between the $\zp$ and the $Z$ are their masses and the $\zp$ couplings to new particle species (DM for example). Additionally, we add a universal scale factor ``$a$" in such a way that when $a=1$, then the $\zp$ couples to the SM fermions equivalently to the SM Z, whereas when $a$ differs from 1 we are accounting for suppressed couplings. In addition to the unsuppressed case ($a=1$), we analyze the case where $a=0.5$. In other words, the $\zp$ couplings to the SM are universally scaled down (relative to the purely sequential case) by a factor of 50\%.

There are many types of DM that can interact with $\zp$ bosons. However, to simplify the discussion we restrict ourselves to DM composed of a single  Dirac fermion, $\chi$. Obviously, the introduction of a $\zp$ boson requires, in principle, the inclusion of additional fermions to cancel the triangle anomalies. Since we focus on the collider and DM phenomenology of the $Z^{\prime}$ portal we will remain agnostic about the specific scenario for anomaly cancellation. For a review of this topic concerning $Z^{\prime}$ models see Ref.~\cite{Preskill:1990fr}. Then, a model-independent approach involves parametrizing the simple tree-level $\zp$ interactions as
\begin{equation}
\label{equation: lagrangian}
\mathcal{L} \supset Z^\prime_\mu~\big[ \bar{\chi} \gamma^\mu \left( g_{\chi \text{v}} + g_{\chi a} \gamma^5 \right) \chi + a \times \sum_{f \in \text{SM} }\bar{f} \gamma^\mu \left( g_{f \text{v}} + g_{f a} \gamma^5 \right) f \big]
~,
\end{equation}
where the sum is over all quarks and leptons of the SM (including neutrinos). We note that for Majorana DM, the most significant change would be when considering constraints from direct detection since then $g_{\chi \text{v}} = 0$. For Dirac DM,  these vector interactions are strongly constrained by current measurements of spin-independent scattering with nuclei~\cite{Akerib:2013tjd}, as will be discussed in Sec.~\ref{section: directdetection}. To further simplify the parameter space, we will assume $\chi$'s vector and axial interactions with the $\zp$ are of equal magnitude and in particular that $g_{\chi \text{v}} = g_{\chi a}$. For this reason, we will often refer to the $\zp -$DM coupling simply as $g_\chi \equiv g_{\chi \text{v}} = g_{\chi a}$. Choosing instead $g_{\chi \text{v}} = - g_{\chi a}$ would not significantly change our results. Moreover, if one departs from $g_{\chi v} \sim g_{\chi a}$ couplings assumption, in particular in the scenario which vector couplings are suppressed along with direct detection limits, the collider bounds we derive further will remain at the TeV scale, keeping our overall conclusions unchanged. 

%%%%%%%%%%%%%%%%%%%%%%%%%%%%%%%%%%%%%%%%%%%%%%%%%%%%%%%%%%%
%\subsection{Sequential-like Z$^{\prime}$} 
%\label{section: sequential}

In the case of a purely sequential $\zp$ ($a=1$), the couplings of $\zp$ to the SM fermions are exactly the same as the SM $Z$. Even though this type of $\zp$ is only expected from gauge theories where the $\zp$ and the SM $Z$ have different couplings to new exotic fermions, it is a useful reference model and is analogous to new states of the SM $Z$ in theories of extra dimensions at the weak scale. However, we will additionally consider the parameter space where the couplings $g_{f \text{v}}$, $g_{fa}$ are all universally suppressed by some constant factor ``$a$" introduced in Eq.~\ref{equation: lagrangian}~\cite{Alves:2013tqa}. More precisely, for the $\zp$ couplings to SM quarks and leptons, we set
\begin{align}
g_{u \text{v}} &= \frac{-e}{4} \left( \frac{5}{3} \tan{\theta_w} - \cot{\theta_w} \right) &&,  & g_{u a} &=  \frac{-e}{4} \left( \tan{\theta_w} + \cot{\theta_w} \right)
\nonumber \\
g_{d \text{v}} &= \frac{e}{4} \left( \frac{1}{3} \tan{\theta_w} - \cot{\theta_w} \right)  &&,  & g_{d a} &=  \frac{e}{4} \left( \tan{\theta_w} + \cot{\theta_w} \right)
\nonumber \\
g_{l \text{v}} &= \frac{e}{4} \left( 3 \tan{\theta_w} - \cot{\theta_w} \right)  &&, & g_{la} &=  \frac{e}{4} \left( \tan{\theta_w} + \cot{\theta_w} \right)
\nonumber \\
g_{\nu \text{v}} &=  \frac{e}{4} \left( \tan{\theta_w} + \cot{\theta_w} \right) &&,  & g_{\nu a} &=  \frac{-e}{4} \left( \tan{\theta_w} + \cot{\theta_w} \right)
~,
\end{align}
where $u, d, l, \nu$ corresponds to up-type quarks, down-type quarks, charged leptons, and neutrinos, respectively. Furthermore, $e = g_w \sin{\theta_w}$ is the electromagnetic coupling, $\theta_w$ is the Weinberg angle, and $a=1$ corresponds to an exactly sequential $\zp$. Our results will focus on the values $a=1$ and 0.5.

We emphasize that our model independent approach maps onto several models dark matter models such as, but limited to, \cite{Barger:1980dx,Profumo:2013sca1,Nie:2001ti,Wong:1992qa,Martinez:2014lta} 
%%%%%%%%%%%%%%%%%%%%%%%%%%%%%%%%%%%%%%%%%%%%%%%%%%%%%%%%%%%
\begin{figure}[!t]
\centering
\subfigure[\label{fig01}]{\includegraphics[scale=0.5]{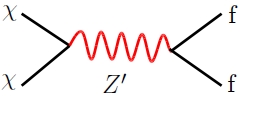}}
\subfigure[\label{fig02}]{\includegraphics[scale=0.5]{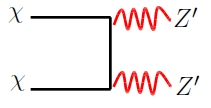}}
\subfigure[\label{fig03}]{\includegraphics[scale=0.5]{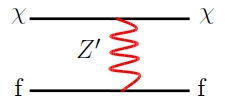}}
\subfigure[\label{fig04}]{\includegraphics[scale=0.5]{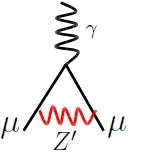}}
\caption{a,b) annihilation channels; c) scattering diagram; d) muon magnetic moment diagram.}
\end{figure}

\section{Direct Detection}
\label{section: directdetection}

Given the simplified model of Eq.~({\ref{equation: lagrangian}}), $\chi$ may scatter off of quarks inside nuclei through the t-channel exchange of a $\zp$. Furthermore, if the $\zp$-$\chi$ interactions are sufficiently large, then these scattering rates are subject to current bounds from direct detection experiments like LUX~\cite{Akerib:2013tjd} and XENON100~\cite{Aprile:2013doa}. In this section, we will give the approximate form for these rates. In the limit of small momentum transfer, $\zp$ exchange can generically lead to spin-independent (SI) scattering with nucleons $n$,
\begin{align}
\label{equation: SI}
\sigma^\text{SI (per nucleon)} &\approx \frac{a^2 \mu^2_{\chi n}}{\pi} \Big[ \frac{Z f_\text{prot} + (A-Z) f_\text{neut}}{A} \Big]^2
\nonumber \\
f_\text{prot} &\equiv \frac{g_{\chi \text{v}}}{m_{\zp}^2} \left( 2g_{u \text{v}} + g_{d \text{v}} \right)
\nonumber \\
f_\text{neut} &\equiv \frac{g_{\chi \text{v}}}{m_{\zp}^2} \left( g_{u \text{v}} + 2 g_{d \text{v}} \right)
\end{align}

and spin-dependent (SD) scattering,
\begin{align}
\label{equation: SD}
\sigma^\text{SD (per neutron)} &\approx \frac{3 a^2 \mu^2_{\chi \text{neut}}}{\pi} \frac{g_{\chi a}^2}{m_{\zp}^4} \Big[ g_{u a}  \Delta_u^\text{neut} +  g_{d a} \left( \Delta_d^\text{neut} + \Delta_s^\text{neut}  \right) \Big]^2
~,
\end{align}
where $\mu_{\chi n}$ is the WIMP-nucleon reduced mass, $Z$, $A$ are the atomic number, atomic mass of the target nucleus, respectively, and $\Delta_q^\text{neut}$ are the quark spin fractions of the neutron. We will take these to be $\Delta_u^\text{neut} = -0.42$, $\Delta_d^\text{neut} = 0.85$, $\Delta_s^\text{neut} = -0.08$~\cite{Cheng:2012qr} and require the spin-independent and spin-dependent rates to be below current published limits from LUX~\cite{Akerib:2013tjd} and XENON100~\cite{Aprile:2013doa}, respectively. Note that for spin-dependent rates, we consider scattering with neutrons since these limits are usually stronger.   

%In Eq.~(\ref{equation: SI})-(\ref{equation: SD}), we have ignored the sub-dominant terms that are suppressed by the incoming momentum of the scattering DM. However, when $g_{\chi \text{v}} g_{q \text{v}}$ or $g_{\chi a} g_{q a}$ vanish, these kinematically suppressed terms become relevant and we may approximate the scattering rates as

%\begin{align}
%\sigma^\text{SI (per nucleon)} &\approx \frac{2}{\pi} \frac{ \mu_{\chi n}^4 ( m_\chi - m_n )^2 v^2}{ m_\chi^2 m_n^2}  \Big[ \frac{Z f_\text{prot} + (A-Z) f_\text{neut}}{A} \Big]^2
%\nonumber \\
%f_\text{prot} &\equiv \frac{g_{\chi a}}{m_{\zp}^2} \left( 2g_{u \text{v}} + g_{d \text{v}} \right)
%\nonumber \\
%f_\text{neut} &\equiv \frac{g_{\chi a}}{m_{\zp}^2} \left( g_{u \text{v}} + 2 g_{d \text{v}} \right)
%~,
%\end{align}

%and

%\begin{align}
%\sigma^\text{SD (per neutron)} \approx \frac{3 \mu_{\chi n}^4  ( m_\chi + m_\text{neut} )^2 v^2}{2 \pi m_\chi^2 m_\text{neut}^2} \frac{g_{\chi \text{v}}^2}{m_{\zp}^4} \Big[ g_{u a}  \Delta_u^\text{neut} +  g_{d a} \left( \Delta_d^\text{neut} + \Delta_s^\text{neut}  \right) \Big]^2
%,
%\end{align}
%where $v \sim 10^{-3}$ is the relative velocity of the incoming scattering $\chi$. Further we comment on the DM annihilation cross section and how we compute the indirect detection limits.

%%%%%%%%%%%%%%%%%%%%%%%%%%%%%%%%%%%%%%%%%%%%%%%%%%%%%%%%%%%
\section{Indirect Detection}
If $\chi$ is thermally populated through the interactions present in Eq.~(\ref{equation: lagrangian}), it is possible that its annihilation rate into SM fermions is still large today. In fact, annihilations into final state quarks and leptons can lead to a diffuse emission of high energy gamma-rays through the processes of neutral pion production and final state radiation, respectively~\cite{Bertone:2004pz}. Indirect detection measurements, such as those performed by the Fermi satellite, seek to observe these emissions in the galactic center (GC) of the Milky Way. Of course, certain astrophysical assumptions must be made (e.g. the nature of the DM density profile), but even conservative ones can lead to significant constraining power in the annihilation rate today, $\langle \sigma v \rangle$~\cite{Hooper:2012sr}. In this section, we briefly describe the analytic forms for these annihilations and the method by which we use the results of Fermi observations to constrain the parameter space of our $\zp$ models.

From the Lagrangian of Eq.~(\ref{equation: lagrangian}), the non-relativistic form for the annihilation cross section into a pair of SM fermions, $f$, through an s-channel $\zp$ (assuming that $m_\chi \gtrsim m_f$) is
\begin{align}
\label{equation: annihilation}
\sigma v \left( \chi \bar{\chi} \to f \bar{f} \right) &\approx \frac{a^2 n_c \sqrt{1-\frac{m_f^2}{m_\chi^2}}}{2 \pi m_{\zp}^4 \left( 4 m_\chi^2 - m_{\zp}^2 \right)^2} \bigg\{ g_{fa}^2 \Big[ 2 g_{\chi \text{v}}^2 m_{\zp}^4 \left(m_\chi^2 - m_f^2 \right) + g_{\chi a}^2 m_f^2 \left( 4 m_\chi^2 - m_{\zp}^2\right)^2 \Big]
\nonumber \\
& + g_{\chi \text{v}}^2 g_{f \text{v}}^2 m_{\zp}^4 \left( 2 m_\chi^2 + m_f^2 \right) \bigg\}
~,
\end{align}
where $v$ is the relative velocity of the annihilating DM pair and $n_c$ is the number of colors of the final state SM fermion. When near resonance, the form of Eq.~(\ref{equation: annihilation}) must be properly modified by including the $\zp$ width, $\Gamma_{\zp}$, which from Eq.~(\ref{equation: lagrangian}) has the form
\begin{align}
\Gamma_{\zp} &= \sum_{f \in \text{SM}}\theta \left( m_{\zp} - 2 m_f \right) \frac{a^2 n_c m_{\zp}}{12 \pi} \sqrt{1 - \frac{4 m_f^2}{m_{\zp}^2}}~\left[ g_{f \text{v}}^2 \left(1 + \frac{2 m_f^2}{m_{\zp}^2}\right) + g_{f a}^2 \left( 1 - \frac{4 m_f^2}{m_{\zp}^2}\right) \right]
\nonumber \\
&+ \theta \left( m_{\zp} - 2 m_\chi \right) \frac{m_{\zp}}{12 \pi} \sqrt{1 - \frac{4 m_\chi^2}{m_{\zp}^2}}~\left[ g_{\chi \text{v}}^2 \left(1 + \frac{2 m_\chi^2}{m_{\zp}^2}\right) + g_{\chi a}^2 \left( 1 - \frac{4 m_\chi^2}{m_{\zp}^2}\right) \right]
~,
\end{align}
where $\theta$ is the unit step function. In Figs.~4-9 we present a light blue excluded region that represents the violation of the perturbative limit $\Gamma_{\zp} / m_{\zp} \leq 0.5$. Obviously, in this region the narrow width approximation fails \cite{Abdallah:2014hon}, and our limits are not applicable. We show such regions for completeness though. Experimental searches for spin 1 bosons at the LHC use as benchmarks the Sequential $\zp$ Model and other similar models, which predict a narrow width for the $\zp$ boson. In other words, those searches assume that a narrow resonance can be described by a Breit-Wigner line-shape when studying the distribution of the invariant mass of the decay products, and this assumption definitely breaks down along with the collider bounds for $\Gamma_{\zp}/M_{\zp} \geq 0.5$ \cite{Abdallah:2014hon}. Since the width of the $\zp$ will largely be controlled by decays into DM pairs, large couplings can still be present if $M_{\zp} < 2 m_{\chi}$. Similarly, the relic abundance calculation also requires a narrow width approximation. Hence, hereafter we will show the excluded  large width region in blue in Figs.~4-9.

The variety of channels increases further if annihilation directly into pairs of on-shell $\zp$ bosons becomes kinematically accessible. In this case, the non-relativistic form for the annihilation cross section (not including the decay of the $\zp$) is
\begin{align}
\label{equation: cascade}
\sigma v \left( \chi \bar{\chi} \to \zp \zp \right) &\approx \frac{1}{16 \pi m_\chi^2 m_{\zp}^2} \left(1 - \frac{m_{\zp}^2}{m_\chi^2}\right)^{3/2} \left(1 - \frac{m_{\zp}^2}{2 m_\chi^2}\right)^{-2}
\nonumber \\
&\times \Big[ 8 g_{\chi \text{v}}^2 g_{\chi a}^2 m_\chi^2 + \left( g_{\chi \text{v}}^4 + g_{\chi a}^4 - 6 g_{\chi \text{v}}^2 g_{\chi a}^2\right) m_{\zp}^2 \Big]
~.
\end{align}
Once produced in this manner, the $\zp$ bosons will decay to pairs of SM fermions. Eq.~(\ref{equation: cascade}) must then be modified by including the appropriate factors of branching fractions for $\zp \to f \bar{f}$. Therefore, when $m_\chi \gtrsim m_{\zp}$, these 2-stage, or \emph{cascade}, annihilations can lead to gamma-rays. Compared to the direct annihilation case, the Lorentz boost between the $\zp$ and SM fermion rest frames generally leads to a widening in the energy range of the gamma-ray spectra for cascade annihilations. 

For direct annihilations, as in Eq.~(\ref{equation: annihilation}), we directly utilize {\tt PPPC4DMID} to calculate the spectrum of gamma-rays~\cite{Cirelli:2010xx}. However, when considering the cascade annihilations of Eq.~(\ref{equation: cascade}), the direct spectra obtained by {\tt PPPC4DMID} must be properly integrated over a finite energy range, corresponding to the range of boosts that the final state fermions are kinematically allowed to obtain~\cite{Mardon:2009rc}. 

Given certain astrophysical assumptions, indirect detection measurements can only constrain the quantity $\frac{1}{m_\chi^2}  \sum_f \frac{1}{2} \langle \sigma v \rangle_f N_{\gamma, f}$, where $\langle \sigma v \rangle_f$ and $N_{\gamma,f}$ are the annihilation rate and number of photons emitted (with energy in some specified range) per annihilation for some final state $f$, and the factor of $\frac{1}{2}$ takes into account that Dirac DM is not self-conjugate. In particular, we use the constraints obtained in Table I of~\cite{Hooper:2012sr}, which gives 95\% CL upper bounds on this quantity for different assumptions of the DM profile and gamma-ray energy bin.\footnote{In comparing Table I to Fig.10 of ~\cite{Hooper:2012sr}, we find a discrepancy, which results in the need to rescale the values in Table I by a value of approximately 2.8 (for an NFW profile). This has been confirmed with the authors of ~\cite{Hooper:2012sr}.} In comparing to our theoretical prediction for  $\frac{1}{m_\chi^2}  \sum_f \frac{1}{2} \langle \sigma v \rangle_f N_{\gamma, f}$, we utilize bounds that are derived using a Navarro-Frenk-White (NFW) profile, and in each scenario that we will consider, we use the gamma-ray energy bin that gives the most stringent upper limit in the parameter space of interest. 

In the two previous sections, the direct and indirect detection observables were described. We now briefly review and compute the correction to the muon magnetic moment arising from our model.

%%%%%%%%%%%%%%%%%%%%%%%%%%%%%%%%%%%%%%%%%%%%%%%%%%%%%%%%%%%
\section{The Muon Anomalous Magnetic Moment}

The muon magnetic moment is one of the most well measured observables in particle physics. There is a long standing discrepancy between theory and experiment
of about $3.6\sigma$. The large uncertainty surrounding the theoretical calculation blurs the excess. The current deviation is $\Delta a_{\mu} = 295 \pm 81 \times 10^{-11}$. Out of this $\pm 81\times 10^{-11}$ error, $\pm 51 \times 10^{-11}$ is theoretical, $\pm 39 \times 10^{-11}$ rising from the lowest-order hadronic contribution and  $\pm 26 \times 10^{-11}$ from the hadronic light-by-light correction. Since the $\zp$ interacts with the muon, it also gives rise to corrections to the muon magnetic moment through Fig.~\ref{fig04}, with a contribution found to be \cite{Queiroz:2014zfa,Kelso:2014qka},
\begin{eqnarray}
&&
\Delta a_{\mu} (Z^{\prime}) = \frac{m_\mu^2}{8\pi^2 M_{Z^\prime}^2} \int_0^1 dx \frac{g^2_{\mu \text{v}} P_{\text{v}}(x)+ g^2_{\mu a} P_{a}(x) }{(1-x)(1-\lambda^2 x) +\lambda^2 x},
\label{vectormuon1}
\end{eqnarray} where $\lambda = m_{\mu}/M_{Z^{\prime}}$ and
\begin{eqnarray}
P_{\text{v}}(x) & = & 2 x^2 (1-x) \nonumber\\
P_{a}(x) & = & 2 x(1-x)\cdot (x-4)- 4\lambda^2 \cdot x^3.
\label{vectormuon2}
\end{eqnarray}
In the limit $M_{Z^{\prime}} \gg m_{\mu}$, this integral can be simplified to,
\begin{equation}
\Delta a_{\mu}(Z^{\prime}) = \frac{m_{\mu}^2}{4 \pi^2 M_{Z^\prime}^2}\left(\frac{1}{3}g^2_{\mu \text{v}} - \frac{5}{3}g^2_{\mu a}\right),
\label{vectormuon3}
\end{equation}which agrees with Ref.~\cite{Kelso:2014qka}.
Notice that, depending on the relative values of the vector and axial couplings, the contribution can be either positive or negative. In this work, the contribution is always negative, since the $\zp$ has similar vector and axial couplings to the SM Z. Thus, we can place the $1\sigma$ bounds based on the aforementioned error bar.  In Figs.~3-8, when we refer to the g-2 bound, we will be forcing the $\zp$ contribution to be within  the the $1\sigma$ error bar. 

We now will turn our attention to the collider phenomenology, focusing mostly on the dilepton searches.

%%%%%%%%%%%%%%%%%%%%%%%%%%%%%%%%%%%%%%%%%%%%%%%%%%%%%%%%%%%
\section{Collider Bounds}

\subsection{Bounds from the search for new resonances in dilepton events}

Several extensions of the SM predict the existence of new gauge bosons as a heavy $\zp$, which can be detected by looking for a resonance in dijet or dilepton invariant masses. The most stringent constraints to date on heavy $\zp$ bosons decaying to charged leptons come from the 8 TeV LHC data~\cite{atlaslep}.

We find that, in fact, constraints from dijet and monojet searches for a $\zp$ are competitive only in the leptophobic scenario studied in~\cite{Alves:2013tqa} confirming those results. As long as the couplings of the $\zp$ to charged leptons are non-negligible, the process $pp\to\zp\to \ell^+\ell^+$, $\ell=e,\; \mu$, becomes more relevant compared to dijets and mono-$X$ processes. 

Concerning the comparison between dijet and dilepton processes, as the couplings to leptons and quarks are of electroweak order, the production rates are not drastically different, with quarks being favored due they color multiplicity. However, the backgrounds to dijets involve QCD pair production and are much larger than SM Drell-Yan background in the case of leptons. Hence, monojet searches can be competitive only for large $\zp$ couplings to neutrinos or to DM. However, a scenario with large couplings to neutrinos and small couplings to charged leptons is not as likely. Furthermore, a very large coupling to DM is not favored by direct detection data.
\begin{figure}[!t]
\centering
\includegraphics[scale=0.5]{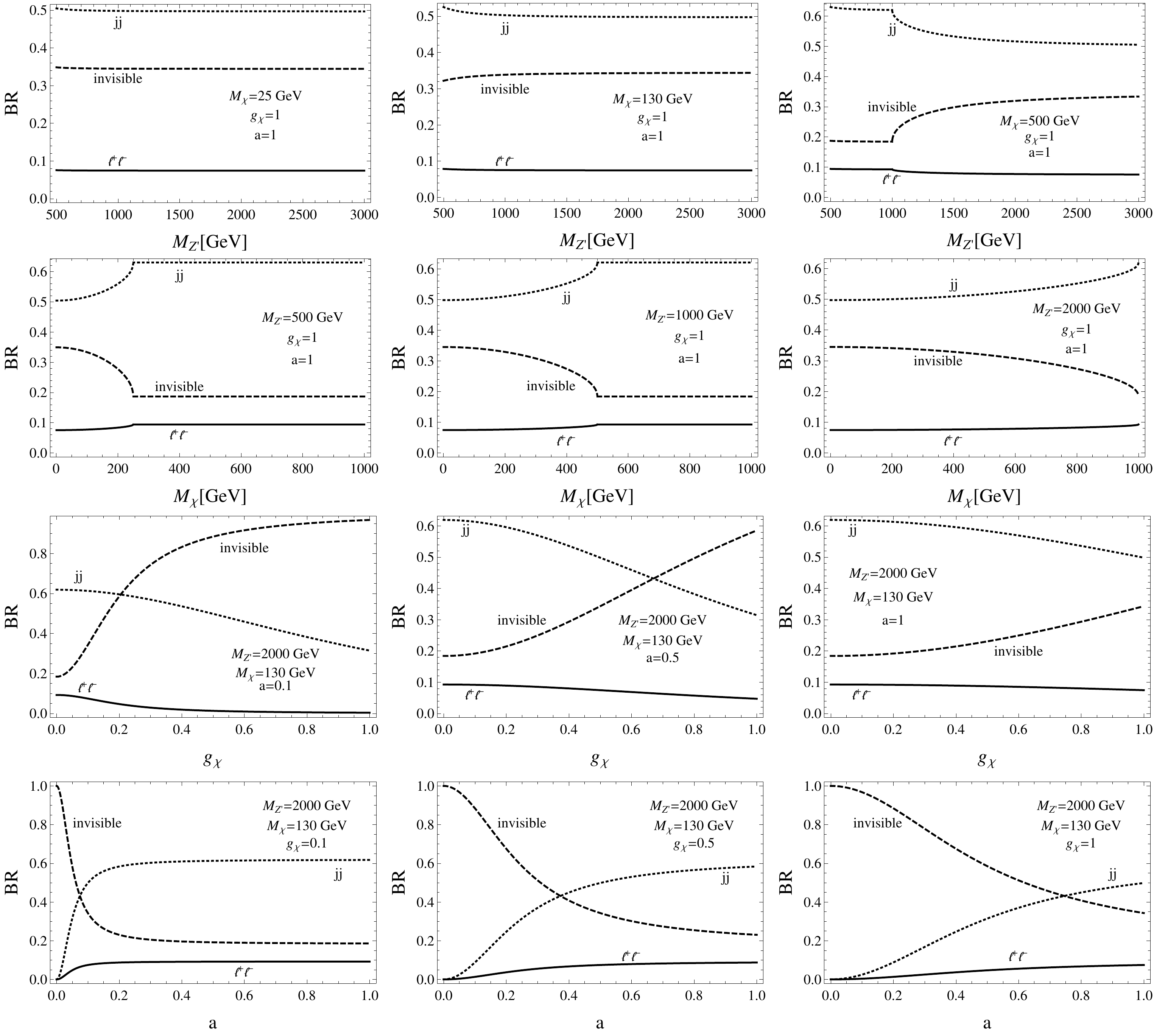}
\caption{Branching ratios of $\zp$ into jets (dotted lines), leptons (solid lines) and an invisible (neutrinos plus DM) mode (dashed lines) as a function of the new gauge boson mass in the first row, the DM mass in the second row, the $Z^\prime$-DM coupling $g_\chi$ in the third row, and the scaling factor $a$ of the coupling between $Z^\prime$ and the SM fermions in the fourth row.}
\label{figbr}
\end{figure}
The effect due to the mass and couplings of the DM particle is indirect concerning the way a dilepton search might constrain a $\zp$ DM model, but not to the way a specific model might fix the couplings between $\zp$ and the SM fermions. In the first row of Fig.~\ref{figbr} we show the branching ratio of $\zp$ into jets, charged leptons and an invisible mode comprising dark matter decays and neutrino decays. The branching ratios are constant as a function of the $\zp$ mass except near the threshold to the dark matter channel. Note that the invisible mode is always sizable because of the neutrinos.

The situation is similar as far as the variation of the dark matter mass is concerned, as shown in the second row of Fig.~\ref{figbr}. The branching ratios of all modes are constant away from the threshold region and the relative size of the branching ratios are controlled by the couplings $a$ and $g_\chi$. To understand how the couplings affect the branching fractions, we display in the third and the fourth rows the variations against $g_\chi$ and $a$, respectively. We kept the $\zp$ and the dark matter masses fixed at a representative point away from any threshold region: $M_\zp=2$ TeV and $M_\chi=130$ GeV. It is clear from these plots that the fraction of decays into dark matter and SM fermions are determined by the size of the couplings we just mentioned.

To summarize, for a fixed $\zp$ mass, the branching ratios into charged leptons $BR(\zp\to \ell^-\ell^+)$ and into jets $BR(\zp\to jj)$ decrease as soon as the DM channel becomes available. Also, decreasing the coupling $\gx$ between the DM and $\zp$ increases the branching faction into leptons and quarks, the same effect occur when we increase the $a$ coupling. However, the branching ratio to charged leptons are much less sensitive to variations of the models parameters compared to the jets and dark matter channels, varying slightly around the 10\% level as can be seen in all panels of Fig.~\ref{figbr}.
%\begin{equation}
%BR(\zp\to \ell^-\ell^+)+BR(\zp\to q\bar{q})+BR(\zp\to \nu\bar{\nu})+BR(\zp\to\chi\bar{\chi})=1.
%\end{equation}
 %We show in Fig.~\ref{figbrzpgx} the branching ratio into charged leptons of a $\zp$ of fixed mass as a function of the DM mass $\mchi$ and the $\gx$ coupling.

The branching ratios to jets are considerably smaller than those found in the leptophobic scenario of \cite{Alves:2013tqa} in the entire $\zp$ mass range for all dark matter masses. The branching ratio to an invisible final state is enhanced compared to a leptophobic scenario due to the appearance of a neutrino channel. As we see in Fig.~\ref{figbr}, the branching ratio to DM plus neutrinos dominate at some points of the parameter space when $g_\chi\gg a$. This is almost twice the typical branching ratio to an invisible channel we might expect in the leptophobic case when $g_\chi\sim 1$~\cite{Alves:2013tqa}. Yet, the impact on the monojet channel is mild. The very hard cut on missing energy imposed by the LHC studies~\cite{mono8} selects mainly events with DM instead of neutrinos.

In summary, allowing sizable $\zp$ couplings to charged leptons softens the constraints found in the leptophobic case studied in \cite{Alves:2013tqa}. In that work, for $g_\chi\le 1$, $\zp$ masses up to $\sim 2$ TeV for light DM could be discarded based on dijet and monojet searches at the Tevatron and LHC, but those limits get much less stringent as $g_\chi$ decreases. As we are going to show, the limits from searches for a dilepton resonance do not soften in the small $\zp$-DM coupling regime. As we now show, dilepton searches are  more efficient to exclude regions of the parameter space of dark $\zp$ models which present sizable lepton couplings.

\vskip0.5cm
\noindent{\underline{Dileptons at the 8 TeV LHC - The ATLAS search}}
\vskip0.5cm

To evaluate the impact of the 8 TeV LHC search on dilepton resonances after $20.3\ifb$ of integrated luminosity for the electron sample and $20.5\ifb$ for the muon sample~\cite{atlaslep}, we simulate the process
\begin{equation}
p\bar{p}\to \zp\to \ell^-\ell^+\; ,\; \ell=e,\mu
\label{ppxll}
\end{equation}
plus up to two extra jets using \texttt{MadGraph5}~\cite{mad5}--FeynRules~\cite{feynrules}, clustering and hadronizing jets with \texttt{Pythia}~\cite{pythia}, and simulating detector effects with \texttt{Delphes3}~\cite{delphes}. Soft and collinear jets from QCD radiation generated by \texttt{Pythia} are consistently merged with the hard radiation calculated from matrix elements in MLM scheme~\cite{mlm} at appropriate matching scales. We adopt the CTEQ6L parton distribution functions computed at $\mu_F=\mu_R=\mzp$. 

The signal events were then selected with the same criteria adopted in \cite{atlaslep} shown below
\begin{eqnarray}
p_T(e_1) & > & 40\gev\; ,\; p_T(e_2) > 30\gev\; ,\; |\eta_e| < 2.47 \\
p_T(\mu_1) & > & 25\gev\; ,\; p_T(\mu_2) > 25\gev \; ,\; |\eta_\mu| < 2.47 \\
128 \text{ GeV} < & M_{\ell\ell} & < 4000 \gev 
\label{cuts}
\end{eqnarray}
Where $\ell_{1}(\ell_2)$ is the hardest (second hardest) lepton in the event, and $ M_{\ell\ell}$ the invariant mass of the lepton pair. The signal acceptance times efficiency found in our simulations are similar to those presented in~\cite{atlaslep}.

All the backgrounds simulations to $pp\to\zp\to\ell^-\ell^+$ were taken from \cite{atlaslep}. To place limits on a $\zp$ model we calculated a $\chi^2$ statistic at the 95\% confidence level based on $M_{\ell\ell}$ measured in \cite{atlaslep} in 6 invariant mass bins: $110-200$ GeV, $200-400$ GeV, $400-800$ GeV, $800-1200$ GeV, $1200-3000$ GeV, and $3000-4500$ GeV. We show in Fig.~\ref{mll} the number of events for the assumed luminosity in each $\ell^-\ell^+$ invariant mass bin for the total background and signal for various values of $M_\zp$, and the production cross section as a function of $M_\zp$.  

The limits obtained from the dilepton invariant mass depend weakly on the DM mass and softens as $g_\chi$ approaches 1. This can be easily understood reminding that the branching ratio to charged leptons are only moderately dependent upon the relevant parameters $M_\chi$, $g_\chi$ and $a$ as can be seen in Fig.~\ref{figbr}, except in the strong $g_\chi$ regime. Overall, the limits obtained on the $\zp$ mass are close to those found in the experimental study of the ATLAS collaboration~\cite{atlaslep}.

We do not take systematic uncertainties into account in the computation of the limits on the $\zp$ mass. For this reason, the colliders bounds are not likely to be conservative. However, the effects from mis-modelling the shape of the signal and background distributions are not expected to be large due the coarse binning used in the experimental study, that is it, it is not necessary a large number of free parameters to fit the distributions. For example, in~\cite{Hoenig:2014dsa} a detailed analysis involving a bin-by-bin systematic uncertainty is shown to have only a moderate effect in the collider bounds even for a finely binning $M_{\ell\ell}$ distribution. 

%We show in Table~(\ref{tablimits}) the exclusion limits on $M_\zp$, in TeV, for the four models that we are considering here for two couplings regimes: $x=1$ and $x=0.5$ and $g_\chi = 0.5$. 

%
\begin{figure}[t]
\centering
\includegraphics[scale=0.55]{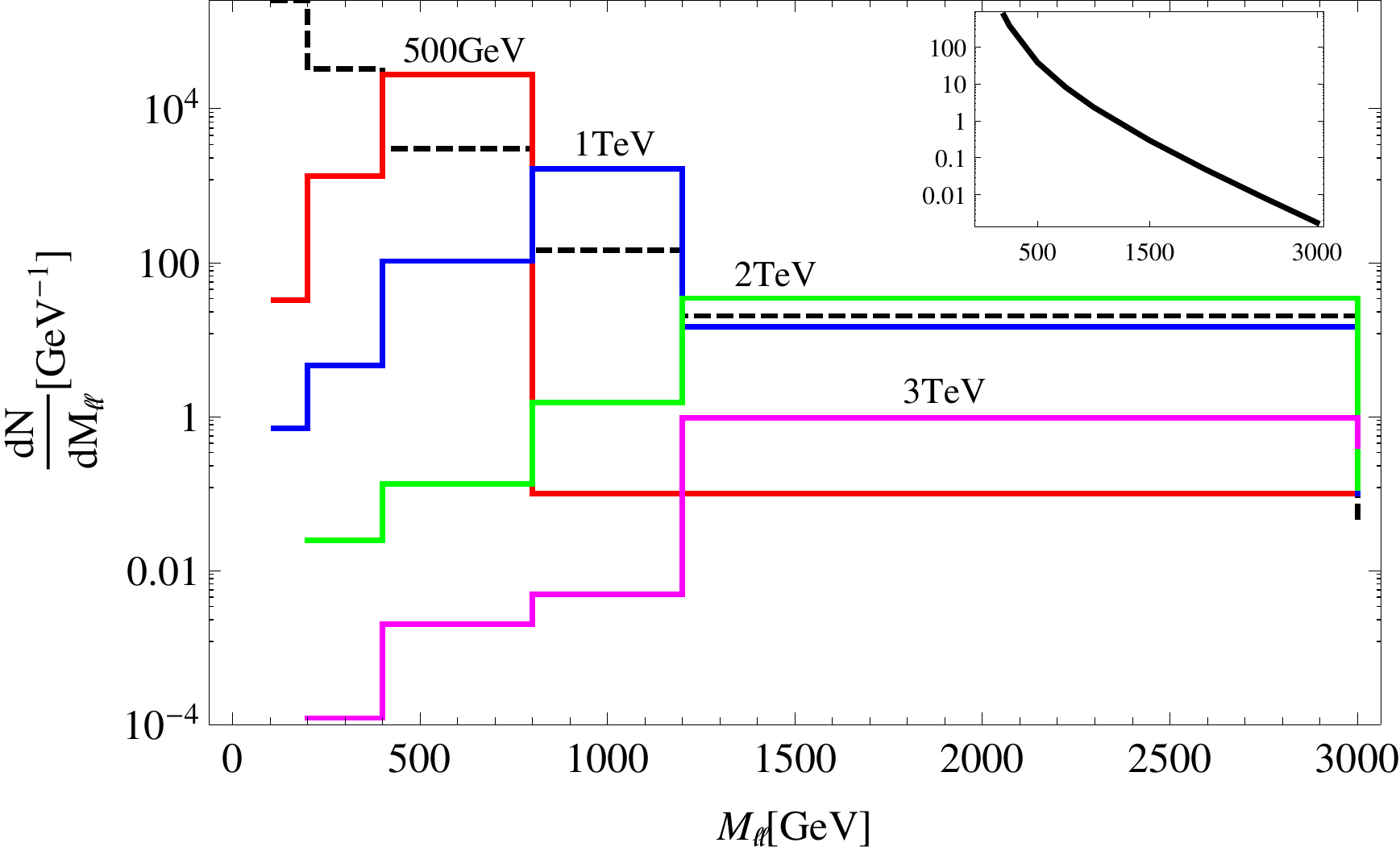}
\caption{The dilepton invariant mass distribution for the total background (dashed line) in six bins taken from \cite{atlaslep} and the signal for four different $\zp$ masses: 500 GeV, 1, 2 and 3 TeV. In the inset plot we show the total cross section in pb as function of $M_{\zp}$. The integrated luminosity assumed in this plot is the same of the experimental study~\cite{atlaslep}.}
\label{mll}
\end{figure}
%

%%%%%%%%%%%%%%%%%%%%%%%%%%%%%%%%%%%%%%%%%%%%%%%%%%%%%%%%%%%
\section{Dark Matter Complementarity}

\begin{figure}[!t]
\centering
\includegraphics[scale=0.45]{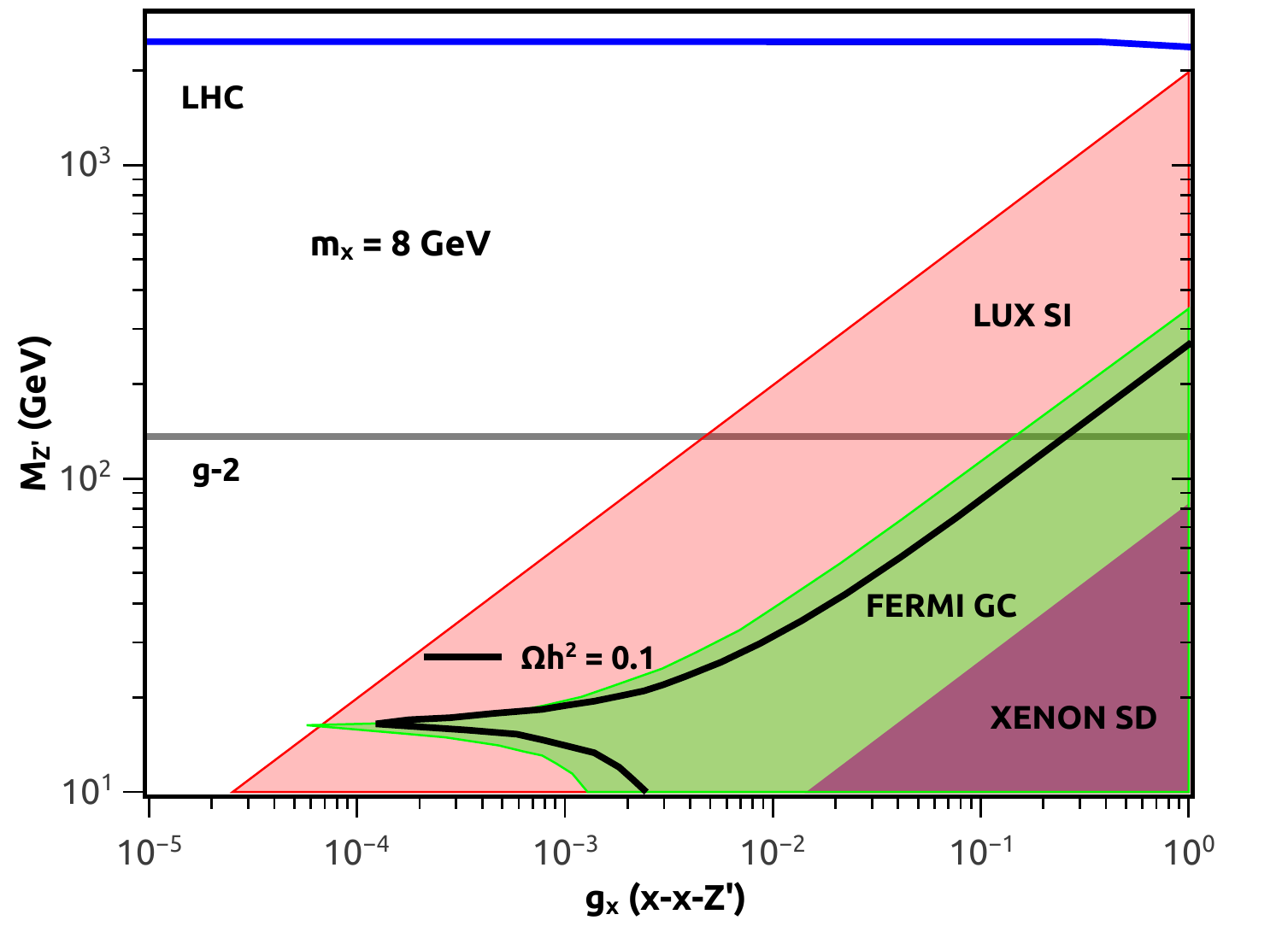}
\includegraphics[scale=0.45]{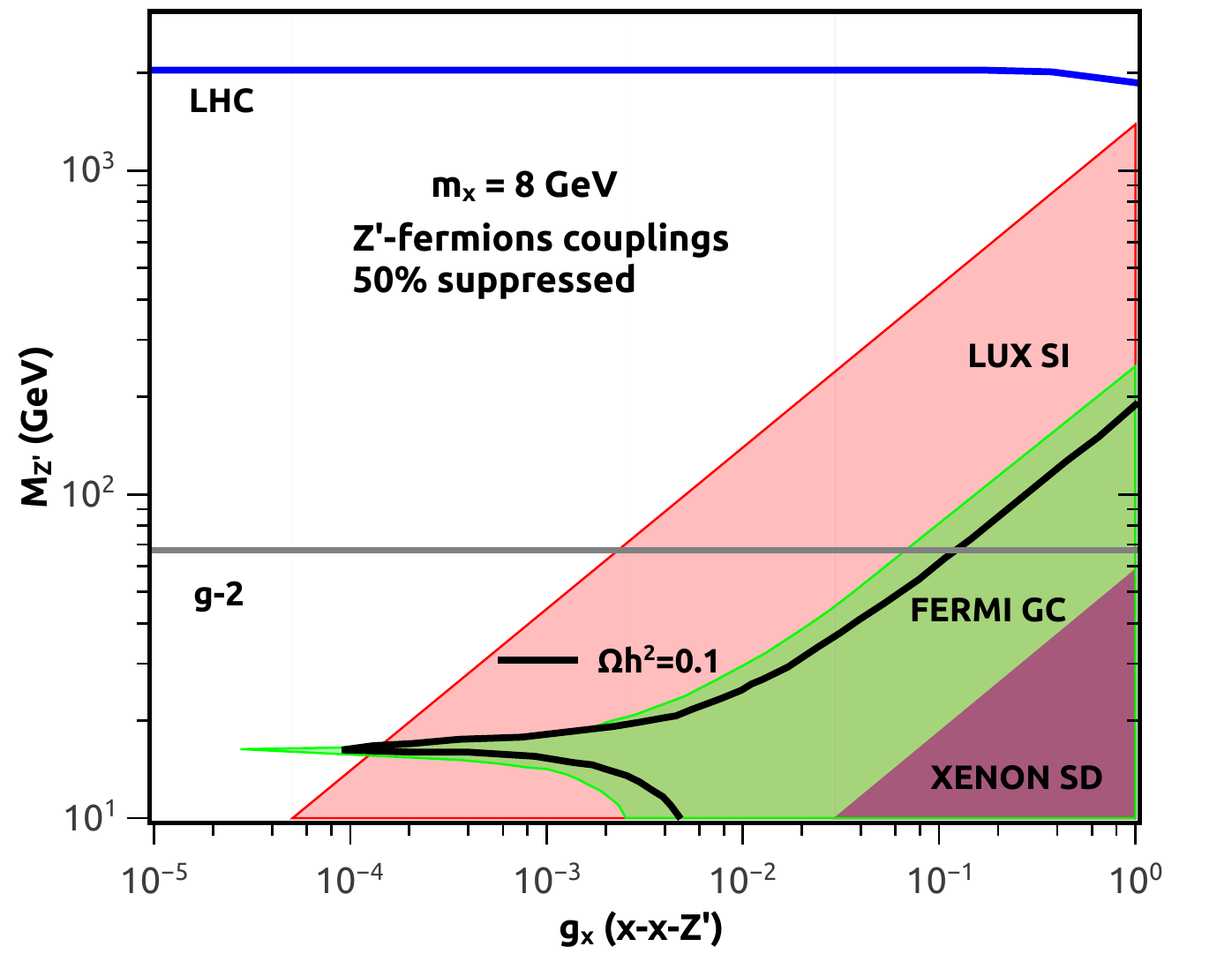}
\caption{Blue horizontal line is LHC exclusion. Everything below the curve is excluded. Gray horizontal line is the $1\sigma$ bound from the g-2. In red, green and pink we show the LUX spin-independent, Fermi Galactic Center, and XENON spin-dependent limits respectively. The black curve yields the right abundance. The light blue region region corresponds to where the perturbative limit on the $\zp$ width is violated. {\it Left:} 8 GeV WIMP with $Z'\equiv Z$;{\it Right:} 8 GeV WIMP with a $Z'$-fermions couplings 50\% suppressed compared with the SM Z. The somewhat wavy aspect of the limit for large couplings occurs due to the coarse binning of the dileptons invariant mass distribution delivered by the experimental study.}
\end{figure}
\begin{figure}[!t]
\centering
\includegraphics[scale=0.45]{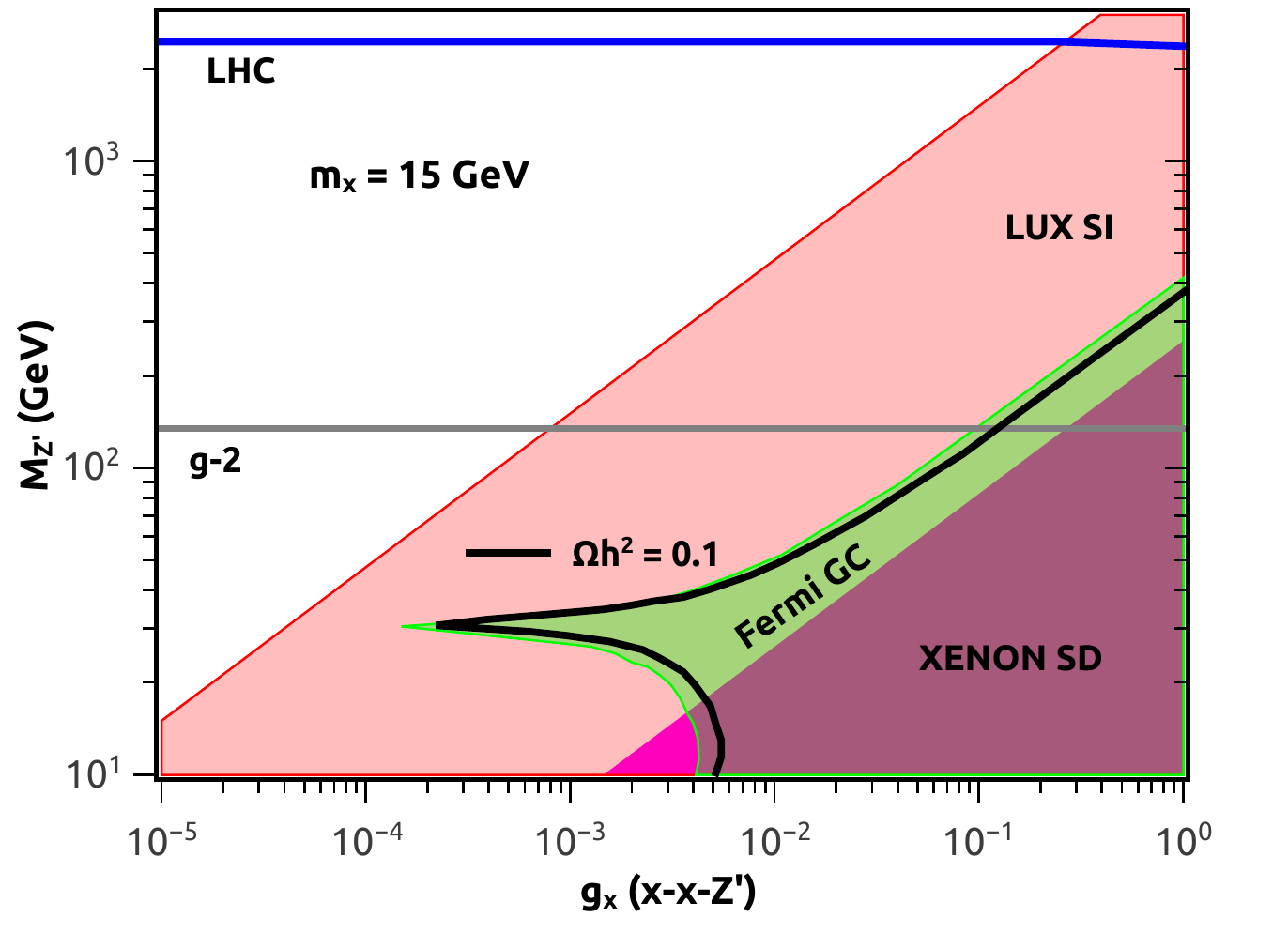}
\includegraphics[scale=0.45]{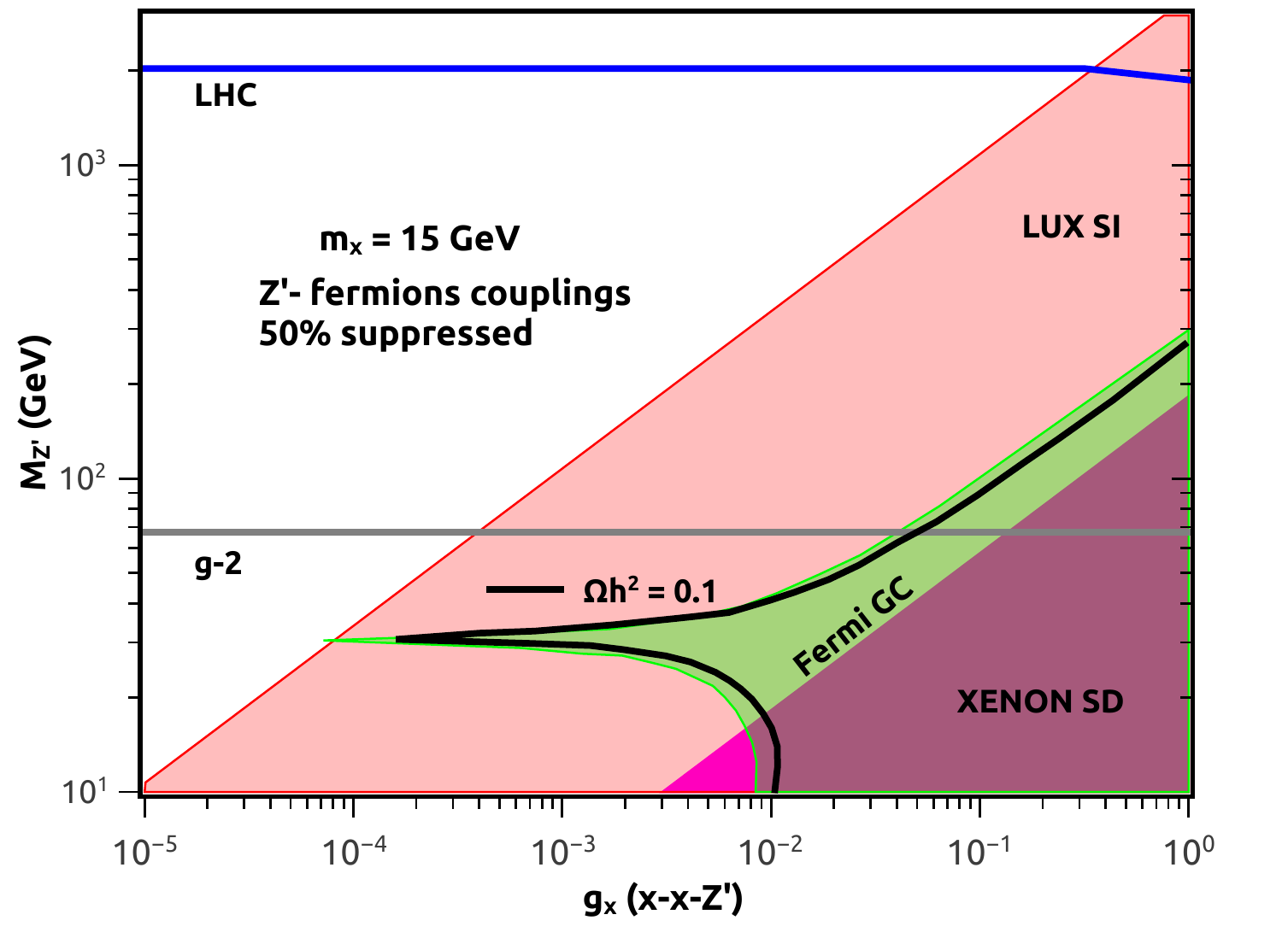}
\caption{We have used the same color pattern throughout. {\it Left:} 15 GeV WIMP with $Z'\equiv Z$; {\it Right:} 15 GeV WIMP with a $Z'$-fermions couplings 50\% suppressed compared with the SM Z.}
\end{figure}
\begin{figure}[!t]
%\centering
\includegraphics[scale=0.5]{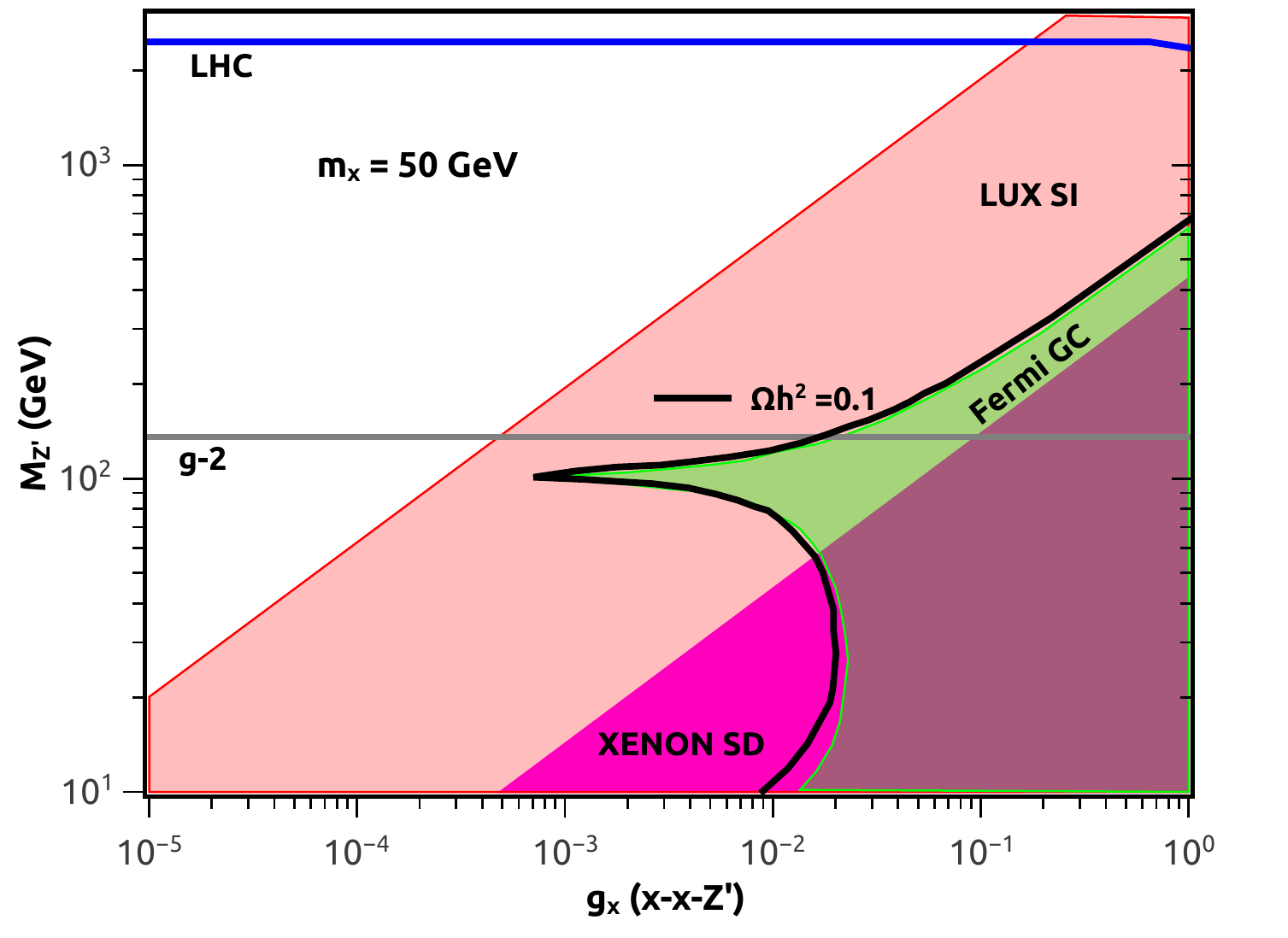}
\includegraphics[scale=0.5]{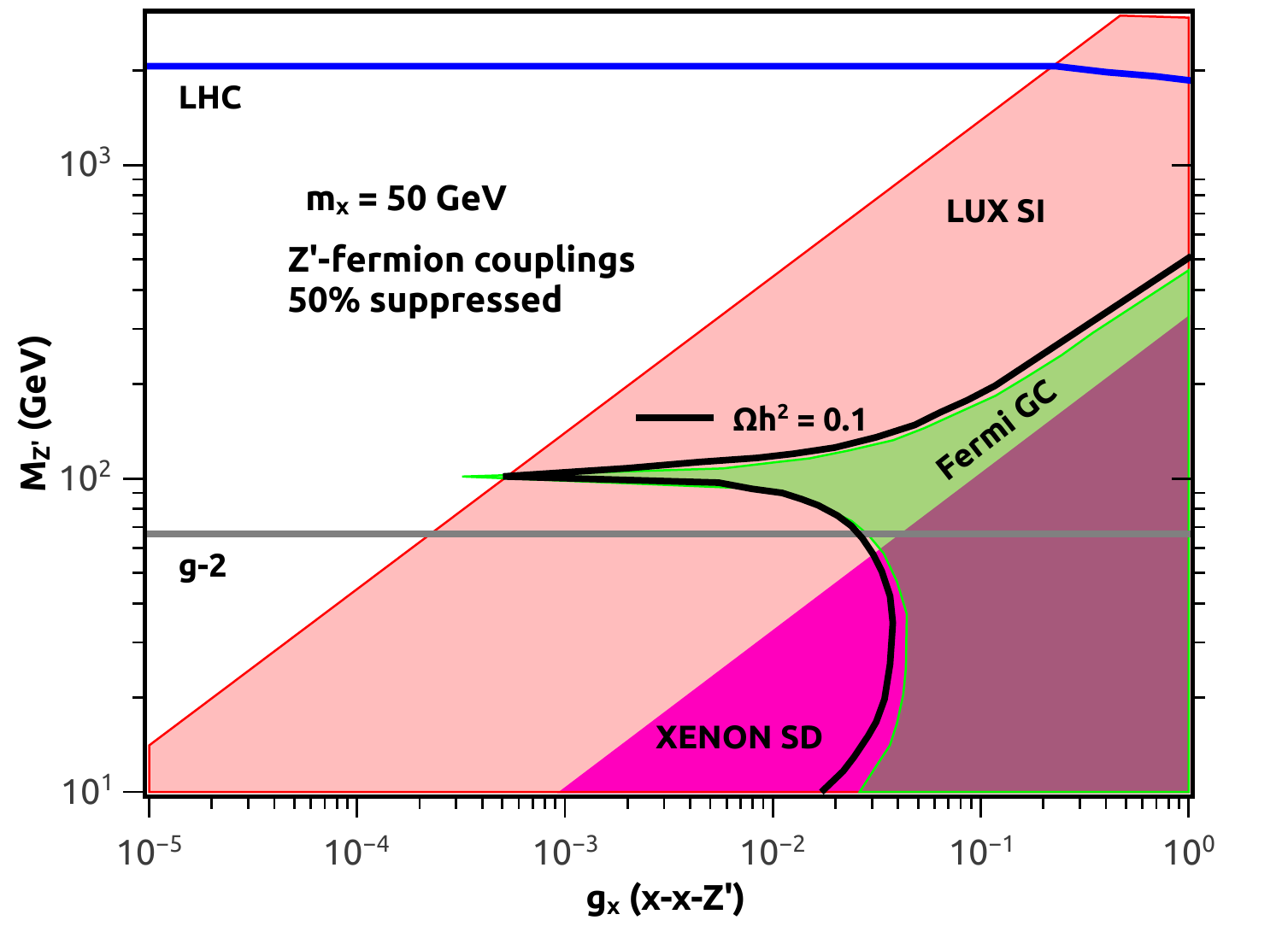}
\caption{{\it Left:} 50 GeV WIMP with $Z'\equiv Z$;{\it Right:} 50 GeV WIMP with a $Z'$-fermions couplings 50\% suppressed compared with the SM Z.}
\end{figure}
\begin{figure}[!t]
%\centering
\includegraphics[scale=0.5]{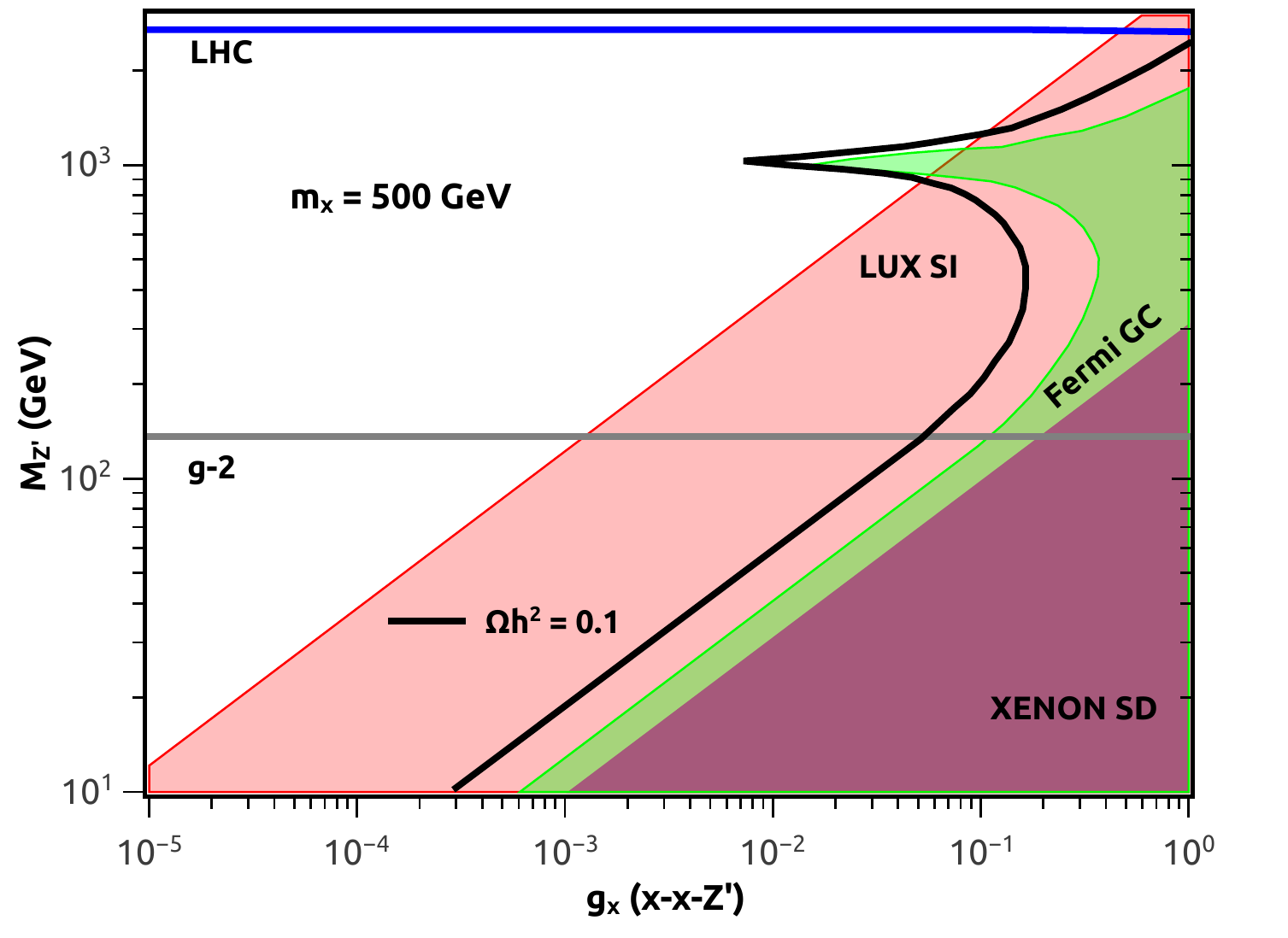}
\includegraphics[scale=0.5]{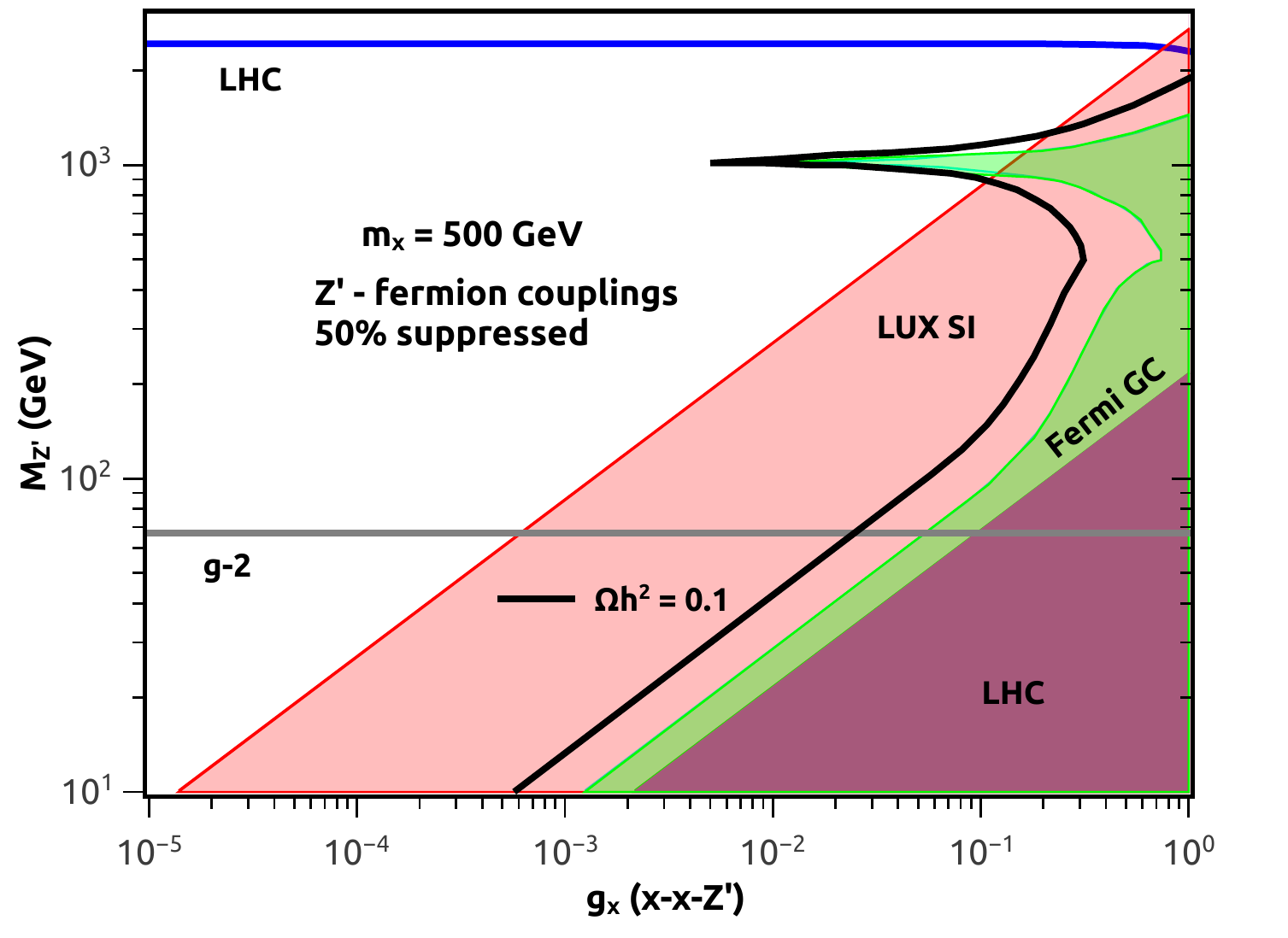}
\caption{{\it Left:} 500 GeV WIMP with $Z'\equiv Z$;{\it Right:} 500 GeV WIMP with a $Z'$-fermions couplings 50\% suppressed compared with the SM Z.}
\end{figure}
\begin{figure}[!t]
%\centering
\includegraphics[scale=0.5]{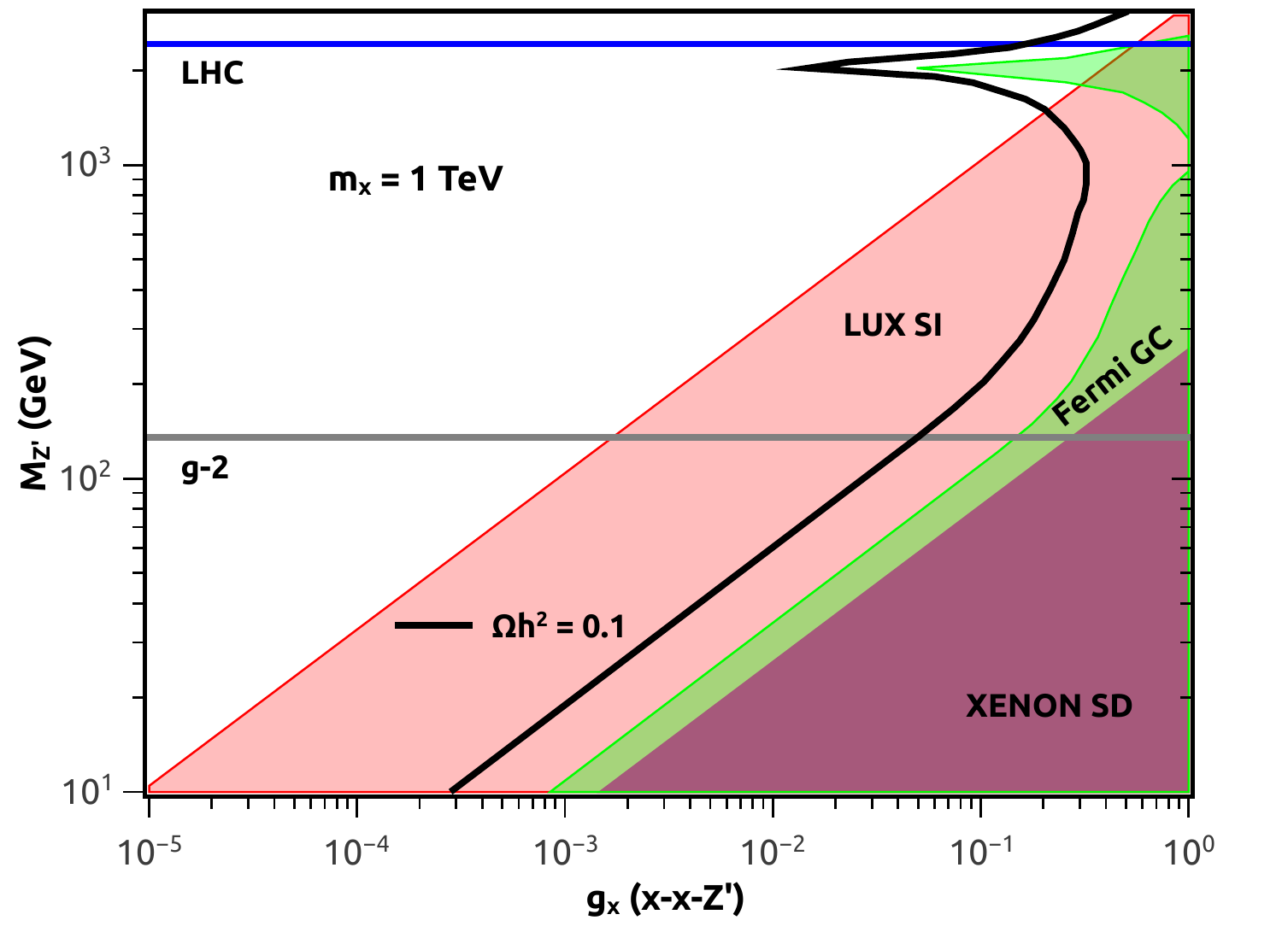}
\includegraphics[scale=0.5]{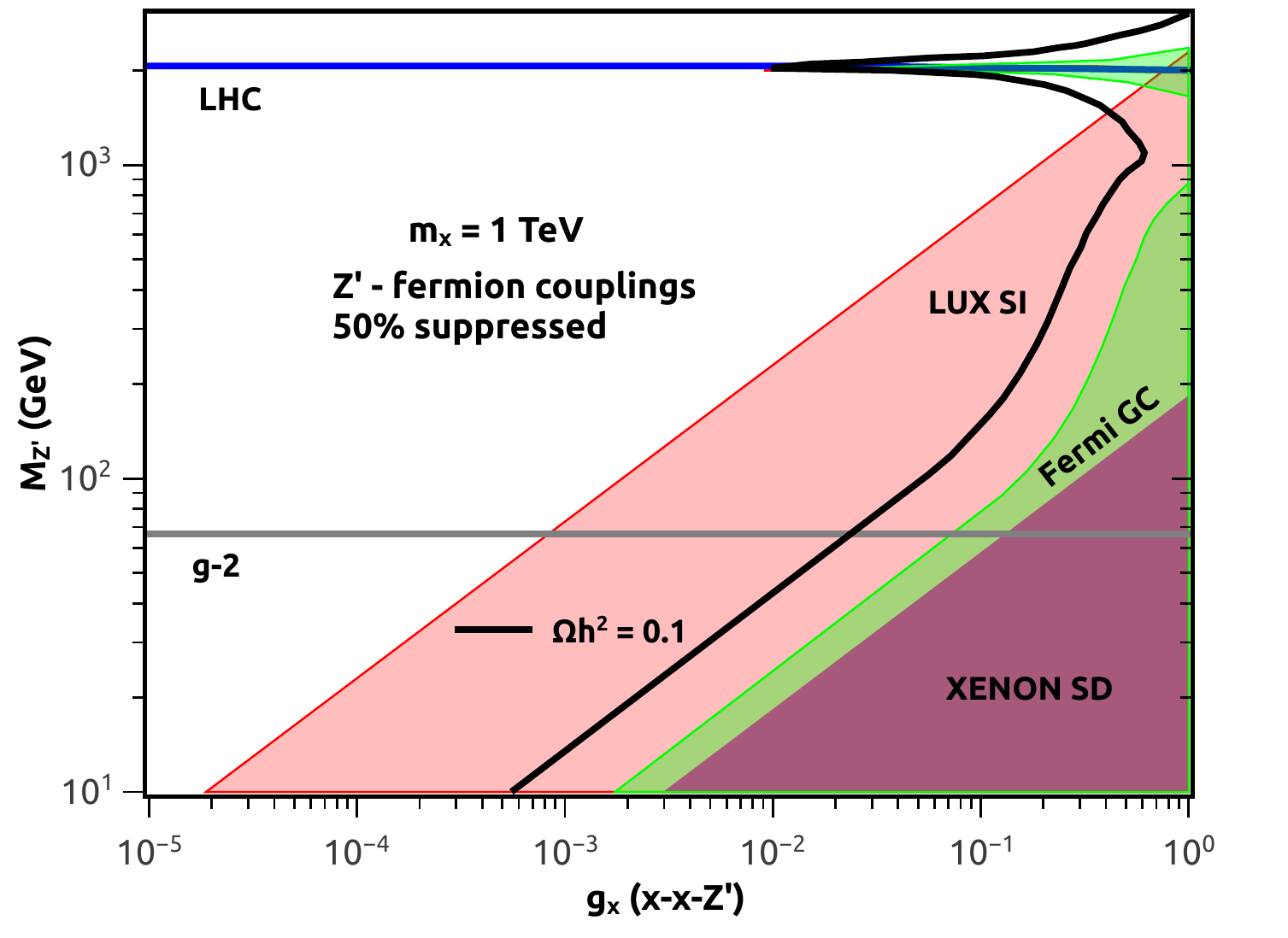}
\caption{{\it Left:} 1 TeV WIMP with $Z'\equiv Z$;{\it Right:} 1 TeV WIMP with a $Z'$-fermions couplings 50\% suppressed compared with the SM Z.}
\end{figure}
\begin{figure}[!th]
%\centering
\includegraphics[scale=0.5]{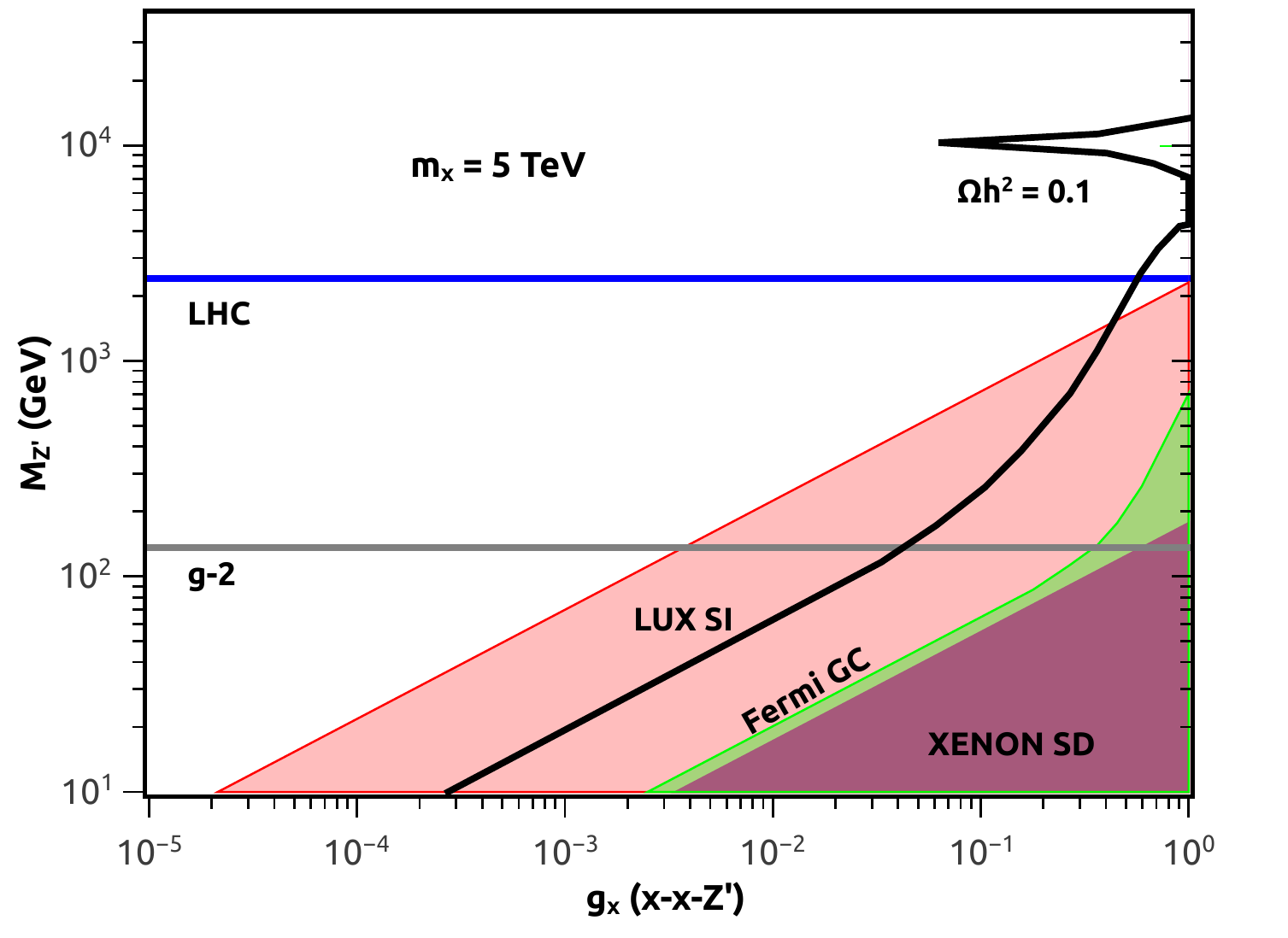}
\includegraphics[scale=0.5]{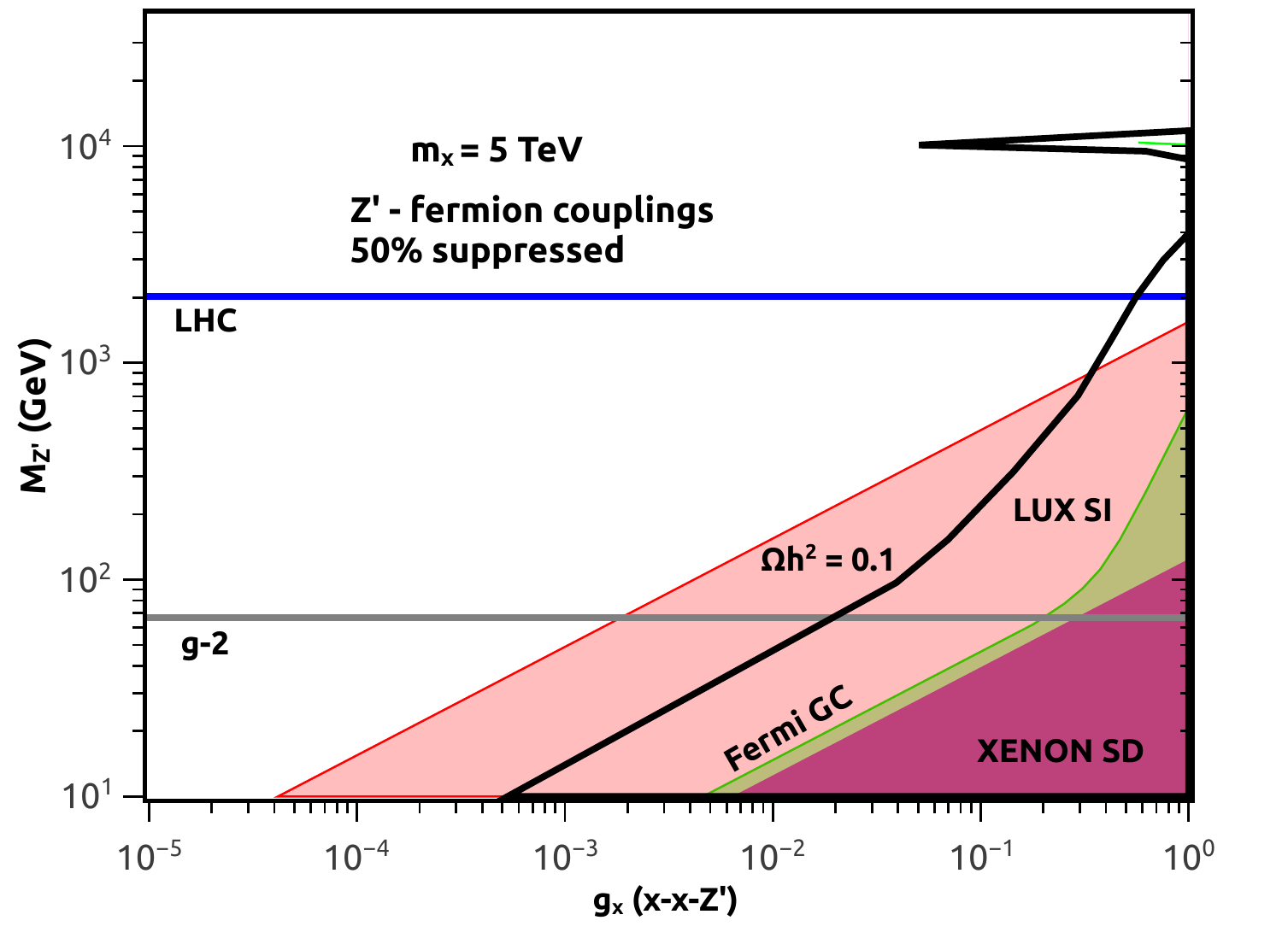}
\caption{{\it Left:} 5 TeV WIMP with $Z'\equiv Z$;{\it Right:} 5 TeV WIMP with a $Z'$-fermions couplings 50\% suppressed compared with the SM Z.}
\end{figure}

We have discussed a collection  of DM constraints stemming from direct, indirect and collider searches and the muon magnetic moment that are relevant for the $Z^{\prime}$ portal scenario under consideration. In this section, we combine these limits and outline the viable {\it versus} excluded region of the parameter space. Also, we explicitly show how critical DM complementarity is. We  focus on  8 GeV, 15 GeV, 50 GeV, 500 GeV, 1 TeV, and 5 TeV WIMP masses, motivated by the recent tentative DM signals and the available LHC data \cite{Daylan:2014rsa}. The direct detection constraints are derived using LUX 2013 and XENON100 2012 results, whereas the indirect detection limits are obtained using the Fermi Galactic Center (Fermi GC) limits from \cite{Hooper:2012sr}. Moreover, the collider limits from dilepton resonance searches are derived using 8 TeV ATLAS data as discussed above. Additionally, we included the perturbative limit on the $\zp$ width, which removes the region with $\Gamma_{\zp}/M_{\zp} \geq 0.5$ \cite{Abdallah:2014hon} as discussed in the previous section. Lastly, the muon magnetic moment limits are obtained using the public code \cite{Queiroz:2014zfa}, where we have simply adapted it to our model. We emphasize that the collider, direct and indirect detection bounds are placed with 95\% C.L, whereas the muon magnetic moment is a $1\sigma$ limit.

In Figs.~4-9 we exhibit the parameter space ruled out by those combined limits, where the importance of DM complementarity is made clear for DM couplings ($g_{\chi}$) larger than unity. 

In the left panel of Fig.~4, we can clearly see even without the inclusion of dilepton bounds that 8 GeV WIMPs are completely excluded by the combined direct detection and the muon magnetic moment limits. Since the $\zp$ is a Z-like gauge boson, the correction to g-2 is negative and therefore a limit of $M_Z' > 135$~GeV is found using Eq.~(\ref{vectormuon3}). As predicted, the muon magnetic moment places complementary limits in the regime of suppressed $Z^{\prime}-\chi-\chi$ couplings, where direct and indirect detection ones become loose. In addition, dilepton data implies $M_\zp > 2.5$~TeV for $g_{\chi} < 0.2$. For couplings larger than unit, the collider bounds are strongly weaken. 

As $g_\chi$ increases, the branching ratio to leptons drops and the collider bounds from dileptons are evaded for all $\zp$ masses. In the region of $g_\chi$ where those bounds weakens the shape of the curves seen in Fig.4, are somewhat wavy due the coarse binning of the dileptons invariant mass distribution delivered by the experimental study. When the $\zp$ mass suddenly populates another bin there's a fast weakening of the bounds causing a tooth-like curve. This feature is repeated for all DM masses.

In the right panel, we have universally suppressed the $\zp$-fermion couplings by 50\%, i.e we set $a = 0.5$ in Eq.~(\ref{equation: lagrangian}). This suppression does not help much, since the entire region of the parameter space that sets the right abundance is ruled out, but the complementarity between DM and collider searches starts to become relevant for large couplings. In this case g-2 demands $M_Z' > 67$~GeV, and dileptons data requires $M_\zp > 2.1$~TeV. In this regime is clear the importance of exploiting complementary searches.

In Fig.~5, it is already quite noticeable the complementarity among direct, indirect, collider searches, and g-2 mostly for WIMP masses of $15$~GeV as we are now away from the energy threshold in direct detection experiments.  Again, the dileptons constraints are the most stringent ones, ruling out $\zp$ masses below $2.5$~TeV ($2.1$~TeV) in the left (right) panel for couplings smaller than 0.1. For larger couplings the direct detection limits are the most stringent ones. The bounds from the muon magnetic moment will follow the one in Fig.~4 since it is independent of the DM mass.

Similar conclusions are found for $m_{\chi}=50$ and $500$~GeV in Figs.~6-7. In Fig.~8, for $m_{\chi}=1$~TeV, we start seeing non-collider bounds becoming more competitive for large DM couplings and some region of the parameter space that reproduces the correct relic abundance surviving all constraints. Interestingly, the combination of g-2, direct, indirect and collider data excludes $g_\chi < 10^{-1}$ in the left panel and  $g_\chi <10^{-2}$ in the right panel and favors only  heavy mediators. Lastly in Fig.~9 for $m_{\chi} =5$~TeV, we see that this feature from Fig.~8 continues and now a much larger region obeys all limits. Albeit, interestingly the perturbative limit on the $\zp$ width removes the region $g_{\chi} \sim 10$ and $M_{\zp} \sim 10$~TeV. 

In summary, when one properly takes into account dilepton, g-2, direct and indirect detection data, { a $\zp$ lighter than $2.1$~TeV is ruled out, independent of the DM mass. One therefore needs vector mediators heavier than $2.1$~TeV to accommodate a Dirac fermion DM particle. We emphasize that depending on the $\zp$-fermion coupling strength masses much heavier than $2.1$~TeV might be required.} regime As we increase the DM mass, the indirect detection limits become more relevant . In Figs.8-9 the indirect detection limits are some times stronger than the direct detection ones. Moreover, the dilepton limits which are usually the most relevant ones, become secondary as larger $\zp-DM$ couplings are used, showing a strong degree of complementarity between those searches. We emphasize that we have used a $\zp$-fermion parametrization which is similar to the SM Z one. Since we studied different DM masses,$\zp$-fermion couplings and a large range for the $\zp$-dark matter coupling magnitude we have covered several $U(1)_X$ extensions of the SM, in a model independent fashion. It would be possible to alleviate those constraints by advocating the presence of a leptophobic $\zp$ as in \cite{Alves:2013tqa}, or a Majorana DM fermion such as in Ref.~\cite{Matsumoto:2014rxa}, or possibly a pure axial $\zp$-fermion interactions such as in Ref.~\cite{Hooper:2014fda}. 

If one departs from the $g_{\chi v} \sim g_{\chi a}$ couplings assumption, specially in the regime that vector couplings are dwindled, the direct detection limits are automatically suppressed, whereas the collider bounds will remain in the TeV scale, keeping our overall conclusions unchanged. We point out that the DM phenomenology in this suppressed scenario has already been studied in Ref.\cite{Arcadi:2014lta}.

We point out that if one had used a different parametrization scheme for the $\zp$-fermion interactions mild changes are expected and the general statement that $\zp$ portal only allows heavy mediators is still valid. Obviously, a key assumption made throughout this work pertains to the strength of the $\zp$-fermion coupling. Since we have normalized our results in terms of the SM Z coupling strength, one could evade those limits advocating much more suppressed couplings. On the other hand, one could also use larger $\zp$-fermion couplings as long as the $\zp$ is sufficiently heavy in order to avoid the discussed limits.

\section{Conclusions}

Motivated by potentially exciting direct and indirect detection signals, in this work we exploited DM complementarity in the context of the $\zp$ portal using a generic parametrization of the $\zp$-fermion couplings with Dirac DM. We performed a detailed analysis of collider, direct and indirect DM detection data as well as the muon magnetic moment to outline the viable vs. excluded region of the parameter space. 

A high degree of complementarity is observed at several different levels: The muon magnetic moment provides complementary limits in the regime of suppressed $\chi-\chi-\zp$ couplings, since then only the $\zp-\mu$ coupling strength is relevant. Indirect detection limits, on the other hand, trace back the region of the parameter space that sets the right abundance, regardless of the coupling strength, and rules out WIMP masses below 15 GeV (except for some viable parameter space very close to resonance). Since we are using a generic parametrization of the $\zp$-fermion interactions, both spin-independent and spin-dependent scattering exists in this model, but we have shown in Figs.~4-9 that the LUX spin-independent limits overwhelms the spin-dependent ones due to the well known $A^2$ enhancement from coherent nucleon scattering. Furthermore, we have shown that for $\zp-DM$ couplings smaller than unit, the  dilepton bounds are the most stringent constraints derived in this work, whereas for larger couplings the indirect and most often the direct detection ones are the most restrictive, emphasizing the importance of DM complementarity. They almost exclude the entire region of the parameter space of the model that sets the right abundance allowing only heavy mediators attached to heavy DM. We emphasize that the light DM regime is ruled not matter how heavy the $\zp$ mass and that our limits would be obviously stronger if larger $\zp$-fermion couplings were used. 

{ In conclusion, after varying the DM mass from 10 GeV up to 5 TeV, and after considering suppressed couplings, we have demonstrated that $\zp$ mediators lighter than $\sim 2.1$~TeV are excluded regardless of the DM particle, and, depending on the $\zp-fermion$ coupling strength, much heavier masses are needed to reproduce the DM thermal relic abundance while avoiding existing limits.}

\section*{Acknowledgements}

A. Alves would like to thank Funda\c{c}\~ao de Amparo \`a Pesquisa de Estado de S\~ao Paulo (FAPESP), grant 2013/22079-8 and Conselho Nacional de Desenvolvimento Cientifico (CNPq) grant 307098/2014-1. SP and FSQ are partly supported by the
US Department of Energy, Contract DE-SC0010107-001. AB is supported by the Kavli Institute for Cosmological Physics at the University of Chicago through grant NSF PHY-1125897. FSQ is also thanks Conselho Nacional de Desenvolvimento Cientifico e Tecnologico (CNPq). FSQ thanks FERMILAB, ARGONNE and North Western University for the hospitality, where this projected was partly carried out. The authors are greatly indebted to Dan Hooper for crucial discussions and comments.  The authors also thank Carlos Wagner, Ian Low, Yann Mambrini and Kuver Sinha for fruitful discussions.

\end{document}